\documentclass[12pt]{article}
 \usepackage{amsfonts}
 \usepackage{amssymb,amsmath}
 \usepackage{graphicx} 

 \newcommand{\crlb}[1]{\label{#1}\\[2pt]}
 
 \newcommand{\crld}[1]{\label{#1}}
 \newcommand{\eela}[1]{\quad\hbox{\scriptsize{#1}}\label{#1}\end{eqnarray}}
 \newcommand{\eelb}[1]{\label{#1}\end{eqnarray}}
 
 \newcommand{\newsecb}[2]{\section{#1}\label{#2}\setcounter{equation}{0}}

 \newcommand{\nolabels} {\def\eel{\eelb}\def\eeql{\eeqlb}  \def\crl{\crlb} 
 \def\newsecl{\newsecb}\def\bibiteml{\bibitem} \def\citel{\cite}\def\labell{\crld}}
\newcommand{\eeqla}[1]{\quad\hbox{\scriptsize{#1}}\label{#1}\end{aligned}\end{equation}}
\newcommand{\eeqlb}[1]{\label{#1}\end{aligned}\end{equation}}

\newcommand\publishversion  {\nolabels\setlength{\textheight}{9in}\setlength
    {\oddsidemargin}{0in} \setlength{\textwidth}{6.7in}\setlength{\topmargin}{-0.2in}}

\def\beq{\begin{equation}\begin{aligned}}		\def\eeq{\end{aligned}\end{equation}}
\def\be{\begin{eqnarray}}  					\def\ee{\end{eqnarray}}		
   \def\bi#1{\begin{itemize}\item[#1]} 	     \def\itm#1{\item[#1]} 	   \def\ei{\end{itemize}} 
   \def\eqn#1{(\ref{#1})}
   	 \def\fn{\footnote}	 

		       \def\del{\delta}        \def\lam{\lambda}  
		               
    		  	\def\Del{\Delta}    
	    		        		     \def\vv{\varphi}     
 	 		     	      	\def\thh{\vartheta} 
	     		          		              	  
\def\W{\Omega}    		  		\def\dd{{\rm d}} 		\def\HH{{\mathcal H}}  
\def\OO{{\mathcal O}} 		     
 	 	   	    
          		\def\ra{\rightarrow}	
 		\def\ket{\rangle}

\def\fract#1#2{{\textstyle\frac{#1}{#2}}}	 	 	
\def\ffract#1#2{\raise .2 em\hbox{$\scriptstyle#1\,$}\kern-.34 em/\kern-.34 em\lower .15 em \hbox{$\scriptstyle\,#2$}}
\def\half{\fract12}

\def\bpmatrix{\begin{pmatrix}} 			\def\epmatrix{\end{pmatrix}}
\def\bmatrix{\begin{matrix}} 			\def\ematrix{\end{matrix}} 
\def\bcenter{\begin{center}}			\def\ecenter{\end{center}}
\def\lowerheightfig#1#2#3{\(\raise-#1\hbox{\includegraphics[height=#2]{#3}}\)}
\def\lowerwidthfig#1#2#3{\(\raise-#1\hbox{\includegraphics[width=#2]{#3}}\)}

\def\intt{{\mathrm{int}}}    \def\mmax{{\mathrm{\,max}}}
 \def\tot{{\mathrm{tot}}} \def\Planck{{\mathrm{Planck}}}
  
 \def\BH{{\mathrm{BH}}} \def\SM{{\mathrm{SM}}}

\def\weglaten#1{}	
 
 \def\uul#1{\ul{\ul{#1}}}   \def\uul#1{\ul{#1}}		 \def\uul#1{#1}

 \publishversion
\begin{document}    

\begin{titlepage}
 \title{The Black Hole Firewall Transformation and Realism in Quantum Mechanics}
 \author{Gerard 't~Hooft}
\date{\normalsize
Faculty of Science,
Department of Physics\\
Institute for Theoretical Physics\\
Princetonplein 5,
3584 CC Utrecht \\
\underline{The Netherlands} \\[5pt]
June 21, 2021}
 \maketitle
\begin{quotation} 
\noindent {\large\textbf{Abstract }} \\[5pt] {\small
A procedure to derive a unitary evolution law for a quantised black hole, has been proposed by the author. The proposal requires that one starts off with the entire Penrose diagram for the eternal black hole as the background metric, after which a procedure was proposed to identify the two asymptotic domains 
of this metric, such that they both refer to the same outside world. In this paper we focus on the need to include time reversal in applying this identification. This forces us to postulate the existence of an `anti-vacuum' state in our world, which is the state where energy density reaches a maximal value. We find that this squares well with the deterministic interpretation of quantum mechanics, according to which quantum Hilbert space is to be regarded as the `vector representation' of a real world. 

One has to understand how to deal with gravity in such considerations. The non-perturbative component of the gravitational force seems to involve cut-and-paste procedures as dynamical features of space and time, of which the re-arrangement of space-time into two connected domains in the Penrose diagram is a primary example. Thus we attempt to obtain new insights in the nature of particle interactions at the Planck scale, as well as  quantum mechanics itself.

In this newer version of the paper an important correction is made concerning the antipodal transformation: applying more insight in the situation, as explained in newer publications by the author,
we now regard region \(II\) of the Penrose diagram as an exact quantum copy of region \(I\), still with time being reversed, but interchanging a region of spacetime with its antipodes is now seen to be incorrect.}
\\[-5pt]\end{quotation}

 \end{titlepage}
	
\newsecl{Introduction}{intro.sec}  \setcounter{page}2
There is general agreement that a theoretical study of black holes in a regime where quantum mechanical effects play a role, is important for a more complete understanding of General Relativity and/or some modifications of this theme, in its relation to quantum mechanics. But here already, parts of the universal agreement end. How do we introduce quantum mechanics in a black hole? According to one doctrine, when subject to quantum mechanics, even large black holes exhibit problems with \emph{information loss}\,\cite{Polch.ref}--\cite{Hawkinginfo.ref} and \emph{firewalls}\cite{firewall.ref},  that can only be understood if one invokes superstring theory and/or AdS/CFT conjectures. However, the result of that is rather sobering: all that can be said about the black hole quantum states, seems to be that these will be `chaotic', which almost puts a premature end to all attempts to understand how to characterise these states. 

In our approach, we try to be more systematic in pushing the principles of General Relativity a bit further, postponing the need for `new Planck scale physics' to be limited to Planck scale black holes only. This supposes that we should be able to constrain ourselves to black holes whose detailed effects on space-time are limited to structures with dimensions much larger than the Planck length. It seems to be obvious that such black holes should be described by standard General Relativity alone. Quantum mechanical effects would be limited to the effects of Hawking particles, which have low energies, and as such have very limited effects on the global space-time structure, at least when regarded over limited stretches of time. In this paper we show that there are other topological aspects of joining together patches of space-time, and these, in turn, will imply important features of the theory of black holes as well as quantum mechanics itself.

At isolated subregions of space and time, which will be smooth and locally flat at the Planck scale, one should be able to employ ordinary quantum field theoretical methods to address all relevant dynamical structures that may arise. The only thing then left to do is find a systematic way to glue these isolated pieces of information together, so that everything will be clear. Beware, this will be a delicate procedure.

It sounds like an obviously correct starting point: if \(A\) is large enough, for black holes with mass\fn{\(M_\BH\) will often be written simply as \(M\). Here, \(A\) is \emph{not} the surface area but just a coefficient.} \hbox{\(M_\BH=A.M_\Planck\)}, and size \hbox{\(R_\BH=2A.L_\Planck\)}, one needs the properties of whatever field theory the Standard Model evolves to, up to energy scale \hbox{\(E_\SM=E_\Planck/A\).} 

This is the regime that we now focus upon.  It should be possible to avoid (Super)string theory here. It is the same regime where Hawking radiation becomes exactly calculable, even though it is weak. To derive the existence of Hawking particles, only standard quantum field theories were needed. Yet, in the expression for the Hawking temperature,
\be k_BT_\mathrm H=\frac{\hbar \,c^3}{8\pi GM_\BH}\, ,\ee
we see that thermodynamics (\(k_B\)), Special and General Relativity (\(c\) and \(G\)), and quantum mechanics (\(\hbar\)) all come together.\fn{A very explicit derivation of the Hawking radiation effect can be found in Ref. \cite{GtHSmatrix.ref}. Issues with a factor 2 are postponed to later publications.} This indicates that, indeed at some stages of our understanding, known theories can be used in combination, to arrive at statements that continue to be valid \emph{at} the Planck scale.

In previous accounts, the author explained how to use the physics of gravitational effects due to particles moving into or out of horizons, to derive dynamical laws for the quantum black hole. Now here, we wish to emphasise that \emph{transformations} should be used in order to arrive at more precise, generic results\fn{In previous papers we avoided the use of transformations relating different observers, because at some points, the time coordinate would have to switch sign. The resulting minus signs were difficult to handle, but in the present paper we believe to have this right. The price paid will be some subtle changes in the definitions of particles and antiparticles, see Sections \ref{penrose.sec} and \ref{package.sec}.}. We wish to consider the \emph{quantum data}, at specific isolated points in space and time, as seen by different `observers'. The observers do not need to `measure' anything, but rather `describe' what they see and experience, in a language that can be understood locally. From a philosophical point of view, this should be the thing to demand: the local language is the best one to use for a local description of the events, so we must be clear about what these local data may say.

It is here that, if one follows the older text books and lecture courses literally, one hits upon the black hole information problem and the firewall problem.  According to several authors, these problems imply that the program just sketched above is bound to fail completely. The present author always maintained that this does not need to be so, but it would be a mistake to think that closing one's eyes for the problems is the appropriate solution. To the contrary, we have to look more carefully at what we are doing.

Suppose that there exists a Hilbert space for all quantum states a black hole can be in. According to our views on the interpretation of quantum mechanics\,\cite{GtHCA.ref}, one has to distinguish \emph{realistic} states from \emph{quantum superpositions} of realistic states. The mathematical background is summarised in Appendix \ref{realQM.ap}. The realistic\fn{In this paper, we use the words `realistic' and `deterministic' interchangeably.} states are the states the system can be in when we imagine that these states describe, as accurately as possible, what the most probable truly existing configuration may be. We call those states \emph{beables}. Together, the beables form an orthonormal set of vectors in Hilbert space, which can be, but do not have to be, used as a basis. The `quantum superpositions' are all other states we get when we superimpose two or more beables.

In a black hole, the quantum evolution operator will seem to transform beables into quantum superpositions (superimposables). This may sound mysterious, as it is not true in some simpler quantum systems.\fn{It does happen in most conventional quantum models such as a particle in a potential well, see Appendix \ref{realQM.ap}.} What this actually means for realism is that, every beable evolves into one of many different possible future beables, depending on some very fast fluctuating variables\,\cite{GtH-2020.ref}.  As soon as we lose sight of the fast fluctuating variables, the notion of \emph{probability} enters. \uul{A diffusion process, similar to what was described much earlier} \uul{by Nelson\,\cite{Nelson.ref}, takes place. } In our theories for quantum realism, summarised in Appendix \ref{realQM.ap},   this tendency is explained. If we could follow each and every one of the fast oscillating variables, we would see that beables evolve into beables, but as soon as we lose sight of the fast variables, we have to deal with probabilities, and probability amplitudes, in accordance with a genuine Schr\"odinger equation. Explicit models can be constructed for almost any Schr\"odinger equation one would wish to reproduce. In our recent papers\,\cite{GtH-2020.ref,GtHLHV.ref} it is explained how this can be squared with the apparent contradictions with Bell's theorem\,\cite{Bell1964.ref} and the CHSH inequalities\,\cite{CHSH.ref}, by constructing explicit models.

\uul{In short, if we start with any particular} realistic state of a black hole at time \(t=t_1\), then determinism tells us that the exact moment, and further details, of the implosion event that has generated the black hole, are completely fixed, but \emph{only} if one knows exactly also all fast variables. Averaging over the fast variables will smear the implosions over large ranges of the implosion time and all other details, 
\uul{ in terms of} superpositions of all possible events, leading  to expressions to which the usual quantum Schr\"odinger equation applies, in the usual way. In fact, what we get is a picture of the distant past and the far future of the black hole that is time reversible.

\uul{ The emergence of} superimposed states, where the early implosions and the late explosions occur at \uul{superimposed} moments in time, also causes superpositions in what we intended  to use as our background 
\uul{metric, and this} explains why we now avoid the use of a background metric that includes the effects of such events, replacing it by the metric of an  eternal black hole. This formalism will get its justification a posteriori.

Note that time reversibility appears to be totally absent in most scenarios that investigators are finding up to this day.
In our work, time reversibility is a central issue. Only if one applies statistical methods in order to handle thermodynamics, time reversibility gets out of sight.

The doctrine called quantum mechanics is merely the vector space representation of deterministic events, as described also in Ref.\,\cite{GtHCA.ref}. This implies that we can employ quantum mechanics as it usually is, while keeping in mind what it says about reality: the real events taking place follow from studying the beables that maximally overlap with the vector considered. In practice, this is the application of Born's rule.\,\cite{Born.ref} As soon as we are unable to specify sufficiently precisely the fastest variables, we recover the usual uncertainty relations, and the standard interpretation of quantum mechanics then applies.\fn{Note that, if applied with sufficient care, the Born rule can be extended to past events as easily as to future events, a fact that can be used to ensure that our effective Schr\"odinger equation will be time reversible.}

In the next sections, we discuss a couple of technical issues that are absolutely crucial for understanding the theory that we wish to unfold here. We claim that the firewall problem is an important feature but it can be completely handled by using an important transformation, transforming in-going particles into out-going ones and vice versa\,\cite{GtHfirewalltrf.ref}. 
\uul{The effect of this transformation is, that:} \emph{ A firewall due to an in-going particle may be} \uul{\emph{replaced by out-going particles}, and, }mutatis mutandis, out-going particles may be replaced \uul{by in-going ones.}
Expanding energy/momenta of in- and out-going particles in \emph{spherical harmonics} is crucial.

Then, section \ref{package.sec} explains the theory in detail, step by step. We see that it is roughly eight precisely shaped jig-saw pieces that have to be put together. The resulting construction is a package deal that has to be accepted entirely, or rejected. There is no possibility in-between.  Indeed, one finds numerous publications, \uul{in which} it was claimed that firewalls\,\cite{firewall.ref} would implicate non-local features in black hole evolution; \uul{In our work}, this conclusion is \uul{sharpened:}  the only non-locality we arrive at is the \uul{new boundary conditions} at the horizons, which could be called non-local, in a sense.  In section \ref{conc.sec} we arrive at our conclusion: black holes can be understood best if one also accepts a radical attitude towards the interpretation of quantum mechanics. 

Our views on quantum mechanics have been explained elsewhere; they are summarised in Appendix \ref{realQM.ap}. \uul{I  do realise} that many readers will not go along here; they can just ignore this appendix entirely. Although I think they are useful, these insights in what quantum mechanics really is, are not essential for the jig-saw puzzle.

\uul{Technical aspects} of our calculations are to be found in the references, but these overlap a lot; much of what we need will also  be \uul{summarised} in this paper.

The newer version of this paper contains some important corrections concerning antipodal transformations; these do not affect most of our principal conclusions. See our note at the end of Section \ref{penrose.sec}.

\newsecl{The Penrose Diagram}{penrose.sec} 
To describe any physical state the black hole might be in, we proceed just the way we are used to in quantum field theory: take a classical solution for the field equations, and then subsequently describe the quantum states in terms of small excitations, using perturbation expansions.  The first state we consider is the pure Schwarzschild black hole, where all Standard Model fields take their local vacuum values. This will be a solution of the pure Einstein equations without matter. 
It is usually written as
	\be\dd s^2&=&-\Big(c^2-\frac{2GM}{r}\Big)\dd t^2+\frac{\dd r^2}{1-2GM/rc^2}+r^2\dd\W^2\ ,\\[3pt]
	\dd\W^2&\equiv&\dd\thh^2+\sin^2\thh\,\dd\phi^2\ .\eel{Schw.eq}
As argued and motivated in the previous section, we shall not add the effects of imploding matter or evaporating matter -- these will be considered later. 

\begin{figure}[h]\begin{center} 
\includegraphics[width=.75\textwidth]
{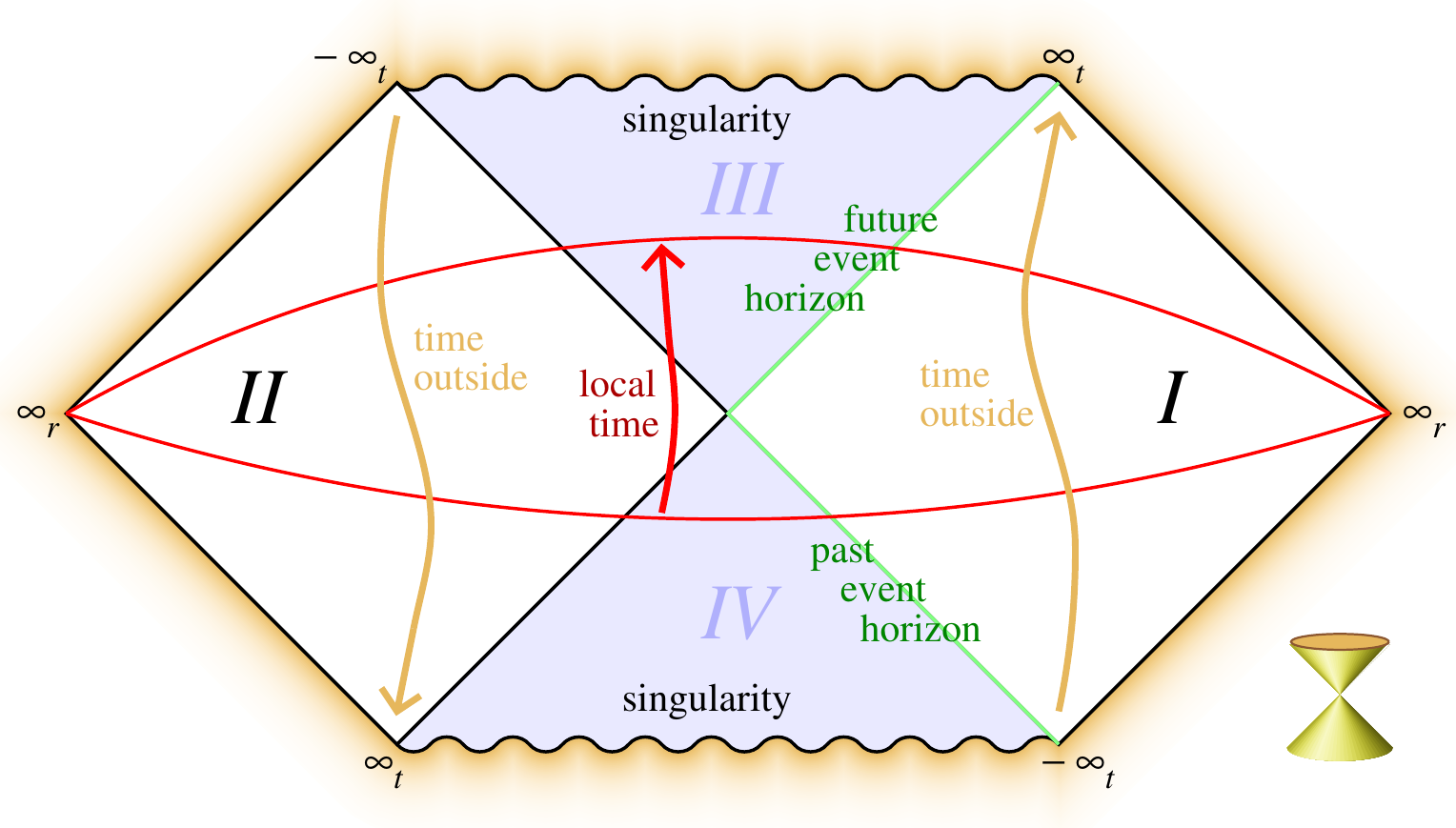}
\end{center}
\begin{quote} 
\vskip-10pt  {\footnotesize\begin{caption}\footnotesize
{Extended metric of a stationary black hole.  The longitudinal coordinates are chosen such that, locally, the light cones are everywhere oriented as indicated in the lower right. Curved lines: Cauchy surfaces; local time runs from below to top, but time for the distant observer reverses in region \(II\) (long arrows).}\labell{penrose.fig}
\end{caption}}\end{quote}\end{figure}

By using different coordinates, one finds that the Schwarzschild solution has an analytic extension beyond the future and the past horizon.  Setting \(c=1\), the Kruskal-Szekeres\,\cite{Kruskal.ref, Szekeres.ref} coordinates \(x\) and \(y\) are defined by
 	\be\Big(\frac{r}{2GM}-1\Big)e^{\,r/2GM}&=&xy\ ;\labell{KS1.eq}\\
			e^{\,t/2GM}&=&x/y\ . \eel{KS2.eq}
In these coordinates,
	\be \dd s^2=\frac{4(2GM)^3}r e^{-r/2GM}\dd x\dd y\ +\ r^2\dd\W^2\ , \eel{ds2KS.eq}
where we see that the horizon singularity at \(r=2GM\) cancels out.

Note that a point \((r,t,\thh,\phi)\) appears\,\fn{In the earliest version of this paper, the angles  \(\thh\) and \(\vv\) in region \(II\) are replaced by their antipodes, \(\pi-\thh\) and \(\vv+\pi\). This is now seen as a mistake, as will be explained.}
 to be mapped onto two points \((x,y,\thh,\phi)\) and \((-x,-y,\thh,\phi)\). 
It is convenient to define coordinates \(\rho\pm\tau=u^\pm\), with
	\be x=\tan(\half\pi u^+)\ ,\quad y=\tan(\half\pi u^-)\ ,\qquad -1<u^\pm<1\ .\ee

\uul{To understand the resulting picture} of the geometry, notice that \(x\) and \(y\), or equivalently, \(u^+\) and \(u^-\), are light cone coordinates.
As time \(t\) increases, \(\tau\) increases if \(x\) and \(y\) are positive. But if we switch the signs of both \(x\) and \(y\), equations \eqn{KS1.eq} and \eqn{KS2.eq} stay the same, whereas \(\tau\) decreases as \(t\) increases. \uul{This sign flip} in the flow of time will be a crucial feature of the theory.
Indeed (remark added in version 2), region \(II\) is a CPT image of region \(I\).

The \emph{Penrose diagram, }Fig.~\ref{penrose.fig}, is obtained if the longitudinal coordinates \(r\) and \(t\) are replaced by  these \(\rho\) and \(\tau\), which include the asymptotic domains of \(r\) and \(t\).

In Fig.~\ref{penrose.fig}, the classically accessible domain is indicated as the diamond shaped region \(I\). \uul{The upward arrow} labelled as "time outside" \uul{is the arrow}  of the Schwarzschild time coordinate \(t\) (or \(\tau\)). The radial Schwarzschild coordinate \(r\) (or \(\rho\))  is oriented in the horizontal direction.

Analytic extension generates three other domains, labelled \(II\), \(III\) and \(IV\). Of these, \(III\) and \(IV\) are of lesser significance.
One might now be tempted to consider the states obtained by just adding a finite number of particles to this  `empty' configuration, but this does not work exactly as one might expect. This is because \emph{time for an outside observer} in region \(II\) runs in the opposite direction, see Fig.~\ref{penrose.fig}: this is where \(x\) and \(y\) are both negative.   Consequently, any particles we might be inclined to add in region \(II\), would contribute \emph{negatively} to the total energy obtained as seen by the outside world. That cannot work correctly. In region \(II\), a local observer would be expected to see particles with negative energies. There are no particles with negative energies for a local observer.

To see exactly what has to be done, we first consider the effect of time boosts in the outside regions \(I\) and \(II\). As explained in the previous section, the Schwarzschild solution is invariant under time boosts, like Einstein's equations themselves. However, the Penrose diagram is not. A time boost for far-away regions in \(I\) and \(II\) corresponds to a Lorentz transformation at the origin of the Penrose diagram. The energy eigenstates for the distant observer are Lorentz transformation eigenstates for a local observer near the origin. For small values of the coordinates \(\rho\) and \(\tau\),  a local Lorentz transformation there is generated by the operator \(L_\mathrm{lor}=\rho\, E-\tau\,p_3\)\,, or in a second quantised system,
	\be L_\mathrm{lor}=\int\dd^3  \vec x\,(\rho\,\HH(\vec x)-\tau\,\mathcal P_3
	(\vec x)\,)\,. \eel{loclor.eq}
We see that, in region \(II\), where \(\rho<0\), the local energy density of particles, generated by the Lorentz transformation \eqn{loclor.eq}, is negative\fn{Using  time translation invariance, take the case \(\tau=0\) to see this.}. One concludes that, if local energies of particles are always positive, then we seem to be talking of negative energy particles as seen by observers in region \(II\).

\begin{figure}[h]\begin{center} 
\includegraphics[width=0.75\textwidth]
		   {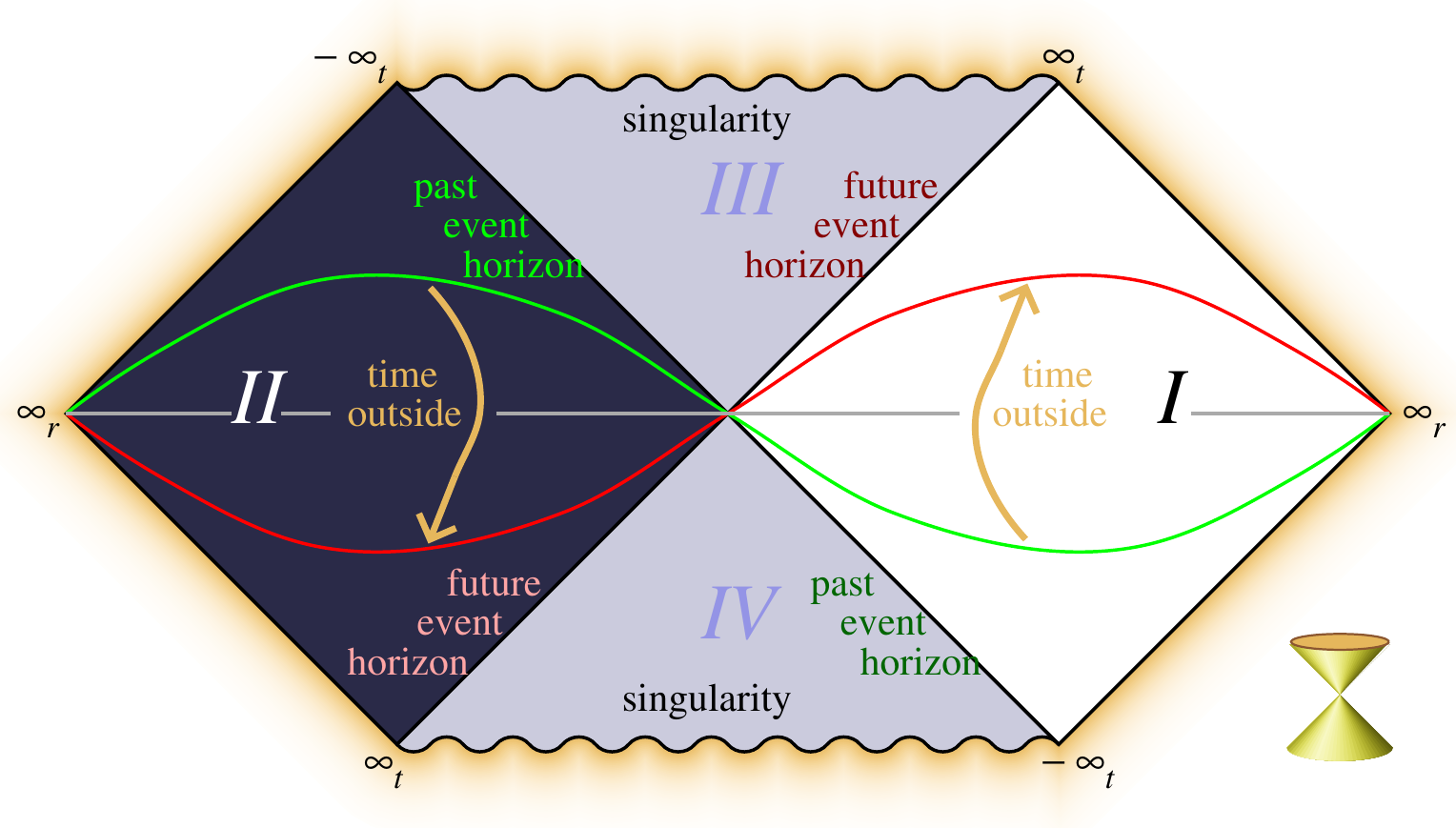}
\end{center}\begin{quote} \vskip-10pt  \footnotesize\begin{caption}\footnotesize
{Penrose diagram with Cauchy surface, at three different times, for distant observers.} Note, that all Cauchy surfaces pivot at the origin. In region \(I\) (white): choose states near the vacuum state, in region \(II\)
(black), states near the antivacuum. \labell{cauchy.fig}\end{caption}\end{quote}\end{figure} 

We now claim that, nevertheless, region \(II\) must be assumed to describe the same universe as region \(I\), since it has an asymptotic region at infinity\fn{Which part of the universe is described by region \(II\)? We discuss this at the end of this Section.}. It would be tempting to side-step this difficulty, but we prefer to face it head-on.  In order to compare physical states in the Penrose diagram with physical states in our universe, we must assume that the particles describing the excited states of the system, not only have a lowest energy state, the vacuum, but also a \emph{maximum} energy state, to be referred to as the `antivacuum'. Then we can simply take region \(II\) to be filled almost to the rim with particles, leaving a few holes, which serve as physical particles in region \(II\). We \emph{shift the definition of energy} for particles in region \(II\), such that their energies take the values \(E^\mmax-\Del E_\mathrm{part}\), where only  \(\Del E_\mathrm{part}\) represents the locally observed energy of a particle in \(II\).
This is exactly the way Dirac introduced the notion of antiparticles. We emphasise this situation in our choice of colors in Fig. \ref{cauchy.fig}.

The reader might wonder how this works out when we add the gravitational forces due to the particles in region \(II\). The answer to this is that we only consider particles (in region \(I\)), or holes (in region \(II\)), with energies that are sufficiently low to allow us to ignore their gravitational back reaction. How this condition works out in detail will become clear soon. As yet, we just declare that gravitational forces will be added as `renormalised' perturbations in terms of gravitons, in order to cancel the effects of \(E^\mmax\).

The time evolution operator is the operator that describes how particles evolve while occupying states on the Cauchy surface. The Cauchy surface is a space-like surface that propagates in the time direction.  For the local observer, \(\rho\) is the space-like coordinate and \(\tau\) is time. In Figure~\ref{penrose.fig}, the Cauchy surface for a \emph{local} observer is shown, at two different values for the time coordinate. The arrow at the center indicates the evolution with time. However, this is not the way a Cauchy surface propagates that describes the data as experienced by a \emph{distant} observer. She uses the Schwarzschild coordinate \(t\) as her time coordinate. Her Cauchy surfaces, at three different times, are indicated in Figure~\ref{cauchy.fig}. Note that these Cauchy surfaces all cross each other at the origin.

Thus, what we found out is how to relate states seen by outside observers, to states observed locally. Cauchy surfaces in region \(II\) must be filled or almost filled with local particles, states in region \(I\) are almost empty. It will be important to realise that, near the origin, `quantum gravity effects' might generate curvatures such that, at very tiny distance scales, the conditions `almost empty' or `almost filled' become ill-defined, so that interesting `quantum gravity' effects are to be expected there.\fn{One could argue what happens in the following way: due to time reversal symmetry, the gravitational effects due to \(E^\mmax\) are to be ignored, but at the boundary between the vacuum and the antivacuum, in a sense, they may be seen to explode. One may say that the boundary between the vacuum and the antivacuum is what remains of the firewall; in Section~\ref{firewalls.sec}, we succeed to take the effects of the firewalls into account completely, leaving almost no trace.}

\uul{We believe that our way of handling the events near the }origin of the Penrose diagram has big advantages as opposed to proposals where some limits to the \emph{acceleration} of particles are proposed,\uul{ see for instance} Rovelli\,\cite {Rovelli}. Such limits require new physics of a kind that is poorly understood. In the formalism here, particles are not accelerated at all, since the background metric is smooth, and the distribution of the physical data on the Cauchy surfaces can be understood directly from flat Minkowski space physics. We do get rapidly varying data near the origin, and these are carefully considered and \uul{handled by}  the firewall transformation, in this paper.

There are still three important issues that need to be considered. First, how can the existence of states with `maximal energy' be reconciled with quantum mechanics? In contrast to standard quantum field theories, the black hole is expected to allow only for a finite number of internal quantum states. Therefore, indeed, one should have expected that there will be a limit to the energies particles in a black hole are allowed to have, so, we should not be surprised to encounter the need for a limit in the allowed energy spectra. Our theory for quantum mechanics has been explained in Refs.\cite{GtHCA.ref,GtH-2020.ref,GtHLHV.ref}. In this approach, the demand for a maximum in the energy appears to be quite welcome. It is associated to a lower limit for the time step variables \(\Del t\) in the classical evolution laws. How this can be squared with Lorentz invariance is not clear, but then, in the Schwarzschild background, Poincar\'e invariance is indeed violated, so, as yet, there seems to be no direct contradiction.

A typical classical dynamical system is as described in our summary of quantum theory in Appendix \ref{realQM.ap}. We find an almost trivial symmetry \(t\leftrightarrow -t\) and associated with that, \(E\leftrightarrow E^\mmax -E\). In the language of the deterministic system, \(E^\mmax\) does not play any other role than being the inverse of the time steps \(\Del t\).
See in particular Eq.~\eqn{discFourier.eq}; the vacuum is at \(k=0\) and the antivacuum at \(k=N-1\), but both these boundaries are totally artificial, they could be displaced in many ways. Fearing that the replacement of the zero energy bound by an \(E^\mmax\) in region \(II\) would distort the gravitational interactions is unfounded. We must however remember to renormalise the gravitational force. Then, it is natural to expect that \(E^\mmax\) cannot generate any anomalous gravitational effects.

The second important issue is the occurrence of two asymptotic regions, one in region \(I\) and one in region \(II\). The outside universe, however, has just a single asymptotic region. This problem does not involve only the time coordinate. Our problem is that there are two universes that are connected in the \uul{spatial} direction.\fn{And we have to keep them connected, in order to be able to execute the firewall transformation, see Section~\ref{firewalls.sec}.}

Consider the one-particle states. A local observer can put the particle at a point in region \(I\) or at the corresponding point, with the same Schwarzschild coordinates, in region \(II\). It looks as if the particle in region \(II\) is a \emph{quantum clone} of the particle in region \(I\). This jeopardises our quantum mapping procedure. We considered to address this problem by dividing 3-space by an isometry subgroup of the form \(\mathbb Z_2\). This reduction of space is as a boundary condition, \uul{since the physical data are now} repeated at the other side of the horizon(s), much like the data of a flat space region when mapped \uul{on a torus.} An important constraint on this boundary condition is that it should not generate any cusp singularities. In the earlier version of this paper the  \emph{antipodal mapping} was proposed to be used as  \(\mathbb Z_2\) in order to avoid this, but we now know that this would generate other difficulties. We now briefly show why the antipodal mapping between regions \(I\) and \(II\) is actually not allowed, in the following 
note added in version 2 of this paper:

In Ref.\,\cite{GtHhydrogen.ref}, the antipodal identification is mentioned explicitly as an elegant solution for the singularity problem at the origin of the Penrose diagram. We now found not only  that the antipodes are not needed, but also that the transition to antipodes\,\cite{Sanchez.ref,Whiting.ref}  for the physical interpretation of region \(II\) does not  obey all our equations.
In Ref.\,\cite{GtHfirewalltrf.ref}, it is explained in more detail how the positions \(u^\pm(\thh,\vv)\) 
and the momenta \(p^\mp(\thh,\vv)\) of the in- and out going particles, can be expanded in spherical harmonics, to 
yield the coefficients
\(u^\pm_{\ell,m}\) and \(p^\mp_{\ell,m}\), all of which now obey evolution and commutation rules
 that are just one-dimensional, since the different components with spherical wave coefficients \(\ell\) and \(m\) do not mix.
Among other results, the commutation rules for these coefficients were found:
\be [u^+_{\ell,m},\,u^-_{\ell',m'}]= i\lam \,\del_{\ell\ell'}\del_{mm'}\ , \qquad\lam=\frac{8\pi G/R^2}{\ell^2+\ell+1}\ . \eel{ellmcomm.eq}
The coefficients \(u^\pm(\thh,\vv)\) must be equal to \(u^\pm(\pi-\thh,\,\vv+\pi)\) at the antipodes, so that
\(u^\pm_{\ell,m}\) are equal to \((-1)^\ell u^\pm_{\ell,m}\) at the antipodes.
The definition of region \(II\) is that this region covers the negative values of \(u^\pm(\thh,\vv)\). Taken together, the 
coefficients \(u^\pm_{\ell,m}\) obey
\be u^\pm_{\ell,m}(-1)^\ell=-u^\pm_{\ell,m}\ , \ee 
so that \(u^\pm_{\ell,m}=0\) if \(\ell\) is even. This violates Eq.\,\eqn{ellmcomm.eq} at even \(\ell\).
Therefore, we have to abandon the antipodal identification for region \(II\). In our later publications, we do have to, and will, deal with the cusp singularity at the origin. End of note.
 
Then, there is a third important issue: the firewalls\,\cite{firewall.ref}. We have chosen to use the Penrose diagram at a given moment \(t_1\) in time
(Sect. \ref{intro.sec}).  At that moment, the Cauchy surface is the grey straight line in Fig~\ref{cauchy.fig}. At times later or earlier than that, the Cauchy surface is distorted, not only as a coordinate effect, but also because of curvature caused by gravitational back reactions. Exactly how to handle this situation appropriately, such as when stationary black holes are considered, where we wish to describe energy eigenstates for the external observer, is discussed in the next section.

\newsecl{The firewall transformation}{firewalls.sec}
In the previous section, it is explained that time translations for a distant observer correspond to Lorentz transformations for the local observer. We also explained that we wish to consider only those particle-like excitations whose energies and momenta are small enough to permit us to neglect temporarily their gravitational back reaction. This is almost a contradiction; we have to sharpen our formulations to include the case that particles are boosted to extraordinary high energies. Here we \emph{cannot} entirely ignore gravity; in contrast, including the gravitational force just here, is possible, and it changes everything; in fact, it becomes the leading effect that dictates how particles will evolve over slightly longer time periods.

This mechanism has been explained in previous publications by the author\cite{GtHfirewalltrf.ref,GtHhydrogen.ref}, but it seems that it still hasn't been appreciated as we think it should. The gravitational force that is responsible here is relatively easy to compute. Since all particles, after strong Lorentz boosts, behave as almost massless objects moving almost at the speed of light, we may assume as a good start,  that the infinite boost limit is appropriate. The precise calculation is in the literature\,\cite{AS.ref,DrayGtH.ref}, but let us first show the hand-waving argument. 

Consider a particle going in. At early time \(t\lesssim t_1\), its energy and momenta are negligible, but when \(t\gg t_1\) their gravitational effect aggravates exponentially. It consists of a shift in the same direction as the particle is going.\fn{We refer to this shift as the `Shapiro effect' or `Shapiro shift', after I.~Shapiro, who first mentioned the gravitational time delay of signals grazing past a heavy object.\cite{Shapiro.ref}} The shift is relatively easy to calculate. Take \(u^-\) as the minus coordinate of some out-going calibration particle. One finds that this particle is shifted by an amount \(\del u^-\) given by
	\be \del u^-=8\pi G f(\W,\W') \,\del p^-\ ; \qquad(1-\Del_\W)f(\W,\W')=\del^2(\W,\W')\ , \eel{shift.eq}
where \(\W\) is the solid angle \((\thh,\vv)\) of the out-going calibration particle, and \(\W'\) the solid angle of the particle going in.
\(\del p^-\) is the momentum of the in-going particle in local coordinates (normalised to the case \(R_\BH=1\)).
The function \(f(\W,\W')\) is a Green function, obeying Eq.~\eqn{shift.eq}, where \(\Del_\W\) is the angular Laplacian. This equation is a linear relation connecting \(\del u^-\) and 
\(\del p^-\), and the fact that it is linear allows us to derive new equations linking in- and out-going particles.

It is very convenient to `renormalise' the equation \eqn{shift.eq}. Instead of regarding it to relate a \emph{shift} \(\del u^-\) to a \emph{modification} \(\del p^-\) of the spectrum of in-going particles, we can re-interpret this as a relation connecting the \emph{average} position \(u^-\) of \emph{all} out-particles, to the \emph{total} momentum \(p^-_\tot\) of \emph{all} particles going in.

These data are dual to one another quantum mechanically, which implies that the entire angular  distribution of all in-going particles is now related to the entire distribution of the out-going ones. This enables us to derive a very good approximation for the \emph{black hole time evolution operator.}\cite{GtHhydrogen.ref}.

This discovery also resolves the problem phrased at the beginning of this section, and this goes as follows. The time-dependence of the momentum of in-going particles is determined by repeated Lorentz boosts. Thus, the momentum increases exponentially with time (we talk of time as defined by an external observer). Consequently, the position of a combination of out-going particles also begins with increasing exponentially with time. Actually, the positions are determined by the local velocity of light -- as long as the out-particles stay relativistic. As the momentum of an out-particle decreases exponentially in time, it sooner or later becomes non-relativistic and continues to obey standard equations, \emph{while moving out} and away from the black hole.

As soon as out-going particles have separated from the black hole, we can remove them from the total picture, at which point also the original responsible in-going particle may be discarded. It was the exponential increase of momenta of in-going particles that caused the \emph{firewalls} there. Thus, we observe that we actually obtained a proper mechanism for removing the firewalls. \emph{A firewall due to an in-going particle may be replaced by  out-going particles. }That's two flies in one blow: the firewall is removed and a possibly quite accurate evolution operator for the black hole is obtained.

This is one of the reasons for regarding our procedure as a package deal. One cannot understand the role of the firewall without understanding the black hole evolution. But it also comes with the other necessary condition: the relation between in- and out-going particles is as the relation between momenta and positions in quantum mechanics: one is the Fourier transform of the other. The Fourier transformation is unitary only if all integrations run from \(-\infty\) to \(+\infty\), not from \(0\) to \(\infty\), which would correspond to keeping all particles in region \(I\). The antipodal identification was proposed in earlier publications, but, as explained at the end of  Section \ref{penrose.sec} in the present version of this paper, we decide to disassociate from it; region \(II\) must be a direct image of region \(I\). The implications of \emph{that} will be further elaborated in a future publication. 

The argument for replacing the firewall with out-going particles can also be time-reversed: the out-going particles are the Hawking particles, which must originate near the horizon where they start off with huge amounts of momentum for each particle. What would the effect of these huge momentum values have been on the metric? It is often claimed that the Hawking particles do not cause a firewall when extrapolated to the past. For some reason, earlier researchers had no problem wishing this firewall away. It just could not be there. Now, we claim that Hawking particles can be regarded as decaying firewalls, in the sense that the mechanism described above works both forward in time as backwards, simply because positions and momenta are quantum mechanically dual to one another.

\newsecl{The black hole theory as a package deal}{package.sec}

We are now in a position to put the various pieces for a sound theory together. What we find from various strains of reasoning, is to be regarded as a jig-saw puzzle. Consisting of eight different jig-saw pieces, it is not too difficult to arrange them in a more transparent structure. First of all, it seems that, when phrasing the set of rules to understand how black holes behave, neither quantum mechanics nor general relativity need any major overhauls, other than considering boundary conditions (such as the firewall transformation) more carefully. Below, we show one result that appears to make sense. The overall formulation is more coherent than what we had before, even if some questions have not yet been answered.

We emphasise that the ensuing black hole theory is a package deal; one has to unwrap all entities in the package, read their manuals, and understand how to deal with them, to see the entire picture.\fn{This situation is not unlike the early days of special relativity, where scientists had to get used to the idea that \emph{time} can not only not be measured independently from the status of an observer, but it cannot even be \emph{defined} independently of observers. Only if one understands that space and time have to be handled as coordinates, which allow for transformations into different coordinates, one begins to understand how special relativity works.}

We have to define a way to characterise all states that together form its Hilbert space, after which we can contemplate the question how these states evolve. At first sight, this seems to be deceptively simple. The situation that we wish to get under control is a black hole with mass \(M_\BH=A M_\Planck\), where \(A\) is sufficiently large, so that it is well distinguishable from the objects that we allow to move in and out, whose typical energies are of the order of \(M_\Planck/A\). These particles move in and out at a sufficiently slow rate to ensure that the black hole stays stationary on time scales considerably longer than \(\OO(A^2 T_\Planck)\).

In that case, we claim that, at first approximation, the description of the quantum states of the black hole, in a given stretch of time\fn{This stretch of time will be kept considerably smaller than the Page time\,\cite{Page.ref}, although the considerations that led Page to define this time period, are of no concern here. The entanglement issues that brought investigators to consider this period are handled differently in the present paper: the basis of Hilbert space must be chosen differently whenever we remove the particles that went too far away to be relevant.}, can be formulated as follows: \\[-20pt]
	\bi{1)} first ignore the gravitational fields of the in- and out-going objects altogether. This means that we will be dealing with the space-time of an eternal black hole, as indicated in Figs.~\ref{penrose.fig} and \ref{cauchy.fig}.
	\itm{2)} This Penrose diagram is not yet suitable for our use because it exhibits two external regions, region \(I\) and region \(II\). We need only one such region, and therefore, as explained at the end of section \ref{penrose.sec}, we divide space-time by a \(\mathbb Z_2\) isometry. In version 1, this was thought to be the antipodal mapping. In future publications, more direct identifications between \(I\) and \(II\) will be considered.
	\itm{3)} Regions \(I\) and \(II\) however must be glued together smoothly, and the gluing prescription must fit snugly over the entire time period. This means that, as seen from region \(I\), region \(II\) seems to be moving backwards in time, implying that, in region \(II\), also the energies are reversed. Ordinary QFT would not allow this, unless we renormalise the energies in region \(II\): given any small subspace entirely embedded inside region \(II\), this subspace must have a \emph{maximum  \(E^\mmax\) for the total energy it can contain,} and if an outside observer sees a particle with energy \(E\) in region \(I\), an observer travelling  in region \(II\), observes an energy \(E^\mmax-E\) in the same subspace. As seen from region \(I\), which may be close to the conventional vacuum, region \(II\) seems to describe it as being almost full; it is close to its antivacuum. As long as we ignore direct gravitational interactions, this is completely consistent.\fn{\uul{How large is} \(E^\mmax\)\,? We can estimate a \uul{reasonable value for} every spherical harmonic \(Y_{\ell m}\). We postulated that the particles considered have energies of order \(E_\Planck/A\). Energies much beyond that value will rarely be relevant. Therefore, we may choose \(E^\mmax \approx E_\Planck /A\). This means that we are doing quantum mechanics using time steps of the order of \(\delta t  \approx A.T_\Planck.\) Remember that each spherical harmonic represents just one physical, particle-like degree of freedom. They cannot be second-quantized. Of course, this argument yields energy and mass values typical for Hawking radiation. Adding all spherical harmonics gives \(E^\mmax_{\mathrm{total}} \approx M_\BH.\)}
	\itm{4)} However, at the point where the two regions border to one another (at \(r\ra 2GM_\BH\), or \(\rho\ra 0\)), the notions of vacuum and antivacuum are ill-defined. We plan not to ignore the gravitational interaction close to the origin, where it matters most. We ignore that only at space-time points sufficiently far separated from the horizons, where all particles have slowed down sufficiently, so that perturbative, renormalised gravity, in combination with Standard Model QFT, is all that is needed to see how these interact.  A valid element of the black hole Hilbert space is now defined by concentrating on one Cauchy surface at fixed time only, say \(t=0\). Left of the origin, where \(x\) and \(y\) are \(<0\), our state rapidly approaches the anti-vacuum. At the right, where \(x>0\) and \(y>0\), the ordinary vacuum is approached. All black hole states are described by specifying how these modes connect at the origin\fn{Remember that, at the origin, transverse space is a projected \(S_2\) sphere whose size does not vanish as in Minkowski space.}. There will always be particles there.
	\itm{5)} The states of the black hole Hilbert space can now be labelled constructively. We must start with one template state; say that this is the state where all fields (except the gravitational metric fields) take their lowest possible energy values in region \(I\) and the highest possible energies in region \(II\). It is not obvious whether this definition would be unique, but in practice, any one state might do as our template state; we go from there to all other states by using creation and annihilation operators. The well-known Bogolyubov transformation\,\cite{Bogoly.ref,Valatin.ref} will tell us how to do such constructions.
	\itm{6)} Consider the addition of one particle with inwards momentum \(p^-\) in region \(I\). Spectator particles moving out will be dragged inwards by the Shapiro effect. As seen by an observer in region \(II\), the spectator particles will be seen to be moved outwards. The rule here is: the entire set of particle data on the Cauchy surface spanned by the past event horizon will be preserved. To be precise: the particle added in region \(I\) is seen to move the state there away from the vacuum state, while in the terminology to be applied in region \(II\), the same particle moves the state there towards the antivacuum. This sign switch matches the sign switch in our definition of energy in the two regions (after the shift with \(E^\mmax\)), and its associated gravitational force.\\
The \emph{firewall transformation} tells us that we can remove the in-going particle if we displace all outgoing particles accordingly. This means that the momentum operators \(p^-(\thh,\vv)\) for the in-going particles can be identified with the position operators \(y(\thh,\vv)\) for the out-going particles, and \emph{mutatis mutandis} we can identify \(-x(\thh,\vv)\) for the in-going particles with the momenta \(p^+(\thh,\vv)\) of the out-going ones. This interchange is to be performed whenever the momentum operators become so large that we can no longer ignore their gravitational back reaction effects.
	\itm{7)} A technical feature that was not yet strongly emphasised in this paper, is that the equations for the gravitational shift (Shapiro shift) that in-going particles bring about in the out-going ones -- and vice versa -- diagonalise if one subjects the in- and out-momenta, as well as the average in- and out positions, to expansions in spherical harmonics.
	\itm{8)} Finally, we have the question how a black hole, with or without an exotic, antipodal, boundary condition, can ever form -- or evaporate completely -- and what the rules are here. This question has not yet been addressed, but we can sketch a reasonably plausible scenario, starting from the conventional description of a black hole being formed by matter. The imploding matter will rapidly form a firewall on the past horizon. A firewall transformation will be needed to replace the imploding matter by Hawking particles going out. The question is then, how the boundary condition is re-generated. How to formulate this elegantly is not yet known;  we had been imagining that first a black hole \emph{seed} is formed, a Planck-sized black hole, close to where the singularity would arise, before we apply the first firewall transformation. This latter point may require  Planck length physics.\ei

\newsecl{Conclusion}{conc.sec}
	As stated in the introduction, one important motivation for the theoretical study of black holes and their quantum features, was to arrive at new insights concerning physics at the Planck scale. 
	
	In our attempts to do things accurately, \uul{one of the first things} noted, is the reversal of time where regions \(I\) and \(II\) in the Penrose diagram, are glued together. This is because a time translation for the  external observer generates a Lorentz transformation near the center of the diagram. \uul{Consequently,} if one regards the Penrose diagram after a (long or short) lapse of time, one sees the Cauchy surfaces pivot around the origin, see Figure~\ref{cauchy.fig}.
	
	\uul{It may sound odd} that this time reversal takes place during the entire lifetime of the black hole, which is much longer than the time scale that can be visualised in the diagram. This is because Lorentz transformations form a non-compact group. It also implies that one cannot ignore its consequences. In particular, if one wants to study the stationary modes -- the energy eigenstates -- of the black hole. 
	
	\uul{After a strong Lorentz boost}, one finds that particles may accumulate towards, or separate from, the centre of the diagram. One of our observations is that the accumulation of particles near the origin may -- and should -- be transformed away, a procedure described here as the `firewall transformation'. It replaced these accumulated in-particles by out particles or \emph{vice versa}. The procedure works, but there may be an important side effect. We ignored this in earlier papers, but it seems inevitable that, when distant observers send in particles with positive energies, the energies may switch sign as well, when a local observer sees them appearing in region \(II\). The sign switch in the energy comes about when one wishes to write the mapping from the familiar world around a black hole to the local world surrounding its horizons, by means of a unitary operation. The time-energy relation in the quantum states, expressed by their phases \(e^{-iEt}\) would ruin unitarity if \(t\) changes sign but not \(E\). Particles can make transitions through the wormhole from \(I\) to \(II\) and back (just because of the Shapiro effect), which is why we do not believe that the energy sign flip can be avoided. Total energy must be conserved.
	
	We were forced to speculate on a limiting local energy, \(E^\mmax\), so that the particles in region \(II\) can have positive energies, \(E^\mmax-E\), while energy conservation is maintained.
\(E^\mmax\) can be related to the smallest time step, \(\delta t\), and this is where deterministic quantum mechanics may enter. Deterministic quantum mechanics leaves some freedom in defining what the smallest time step is, so there may be still some room for speculations concerning the value of \(E^\mmax\). The theory is entirely invariant under \(E\leftrightarrow E^\mmax-E\).
	
	The state with energy \(E^\mmax\) will be called the 
 `antivacuum'. The antivacuum in region \(II\) may actually represent all those particles that filled the black hole during its implosion in the distant past. 
	
As yet in a modest way, our work seems to pay off. Quantum mechanics is likely to demand such a thing as an upper limit for the energy on any compact subset of 3-space -- even when disregarding the direct effects of gravity -- leading to the notion of an antivacuum. In elementary exercises involving the nature of quantum theory in general, the notion of antivacuum was hinted at, if only to account for the existence of a smallest time scale for defining local interactions. Limits on the energy density of quantum states near the Planck scale have been speculated about a lot; here we see the necessity of this, although the mathematical details are still not quite clear.

	\uul{How does gravity act in} region \(II\)? The answer should follow in a straightforward way by analysing the transformations from the asymptotic  Minkowski space to the geometry of region \(II\), or even simpler, by transforming \(I\) to \(II\) and back. The doctrine demands that we start by ignoring the gravitational effects of the quantum excitations that are being considered, just because the masses and energies of these quantum particles were systematically assumed to be orders of magnitude smaller than that of the black hole itself. As the first step towards further refinement, one can then add the gravitational force perturbatively as one can add the other Standard Model interactions. The non-perturbative aspects of gravity are most acute in the immediate vicinity of the horizons, and those were taken care of accurately by taking the Shapiro shifts into account: the firewall transformation.
	
	We found that in describing interacting beams of particles at Planckian energies, firewalls may be created and subsequently removed by utilising the firewall transformation. A super-Planckian particle may come with the gravitational analogue of a shock wave. General Relativity will allow us to remove this shock wave by a cut-and-paste procedure combined with a local coordinate transformation, after which particles going one way (inwards or outwards) are replaced by particles going in the opposite direction, the rules here being precisely dictated by standard physics.
	
	Needless to say, there are still many questions left: a \emph{systematic and complete} formalism for the entire set of physical degrees of freedom at Planckian scales, is still missing. Unlike some procedures in string theory and \(M\) theory, we insist on maintaining fundamental aspects of \emph {locality}, which should considerably constrain the amount of freedom we have in constructing our theory. More than half a century ago, physicists had similar difficulties in reconciling quantum mechanics with \emph{special} relativity. The insight of how to do this right became clear gradually: quantum field theory (QFT). This theory became a beautiful synthesis of simplicity and complexity; the theory was simple enough to enable us to categorise all possible scenarios, and sufficiently complex to allow something as intricate as an entire universe with galaxies, stars, planets and life all controlled by its equations.
	
	We can only hope that the eventual synthesis of quantum mechanics with \emph{general} relativity will be equally simple and complex to give us an even more detailed understanding.
	
In the Standard Model of the subatomic particles, the different energy eigenstates of its Hamiltonian appear to have quite distinctively different physical properties. The importance of its interpretation as a vector representation of a theory with local realism is that, all energy eigenstates just differ from one another by a man-made phase coefficient, which does depend on time. The lowest energy state is the vacuum state, which in our new picture appears not at all to differ from the highest energy state, the `antivacuum'. Because of this, it now looks much more natural than before, to treat vacuum and antivacuum as each other's mirror image.

\appendix \newsecl{Quantum mechanics as the vector representation of a real, deterministic theory}{realQM.ap}
The author's approach to viewing quantum mechanics as a deterministic theory in disguise, has been unfolded in several publications\,\cite{GtHCA.ref,GtH-2020.ref,GtHLHV.ref,GtHonto.ref,Wetterich.ref}. 
Consider a classical, deterministic system, which only allows for `states' that are as `real' as the planets in their elliptical orbits. There is nothing yet to remind us of quantum mechanics, except that the physical variables may refer to phenomena at the tiniest distance\,- and time scales. We label the real states by using the Dirac notation -- but here it is meant to be nothing more than just a notation -- :
\be |0\ket,\ |1\ket,\ \cdots\ |n\ket,\ \dots,\quad \hbox {where\ \ } 	n \ \hbox {\ is integer.} \ \eel{beables.eq}
It will then often be convenient to assume a world where many of its variables are almost periodic. `Almost' will mean in practice: the variable returns to one of its previous values regularly, unless something in its direct environment will cause it te deviate from the rule in exceptional situations. What this means is that we can use the assumption of periodicity as an approximation, and worry about the -- rare -- exceptions later\fn{The rare exceptions are to be expressed in terms of  interaction Hamiltonians, \(H^\intt\), which will have only tiny effects on the states considered.}. Thus, let us start with an `unperturbed' world where everything is periodic (though the periods may differ). A variable \(|\psi\ket\)  then obeys the evolution law
	\be |\psi\ket_t=U_{\Del t}\,|\psi\ket_{t-\Del t}\ ,\qquad U_{\Del t}\,|n\ket=|n+1\ket\ ,\quad |N\ket=|0\ket\ . \eel{classevolv.eq}
Since these states refer to `reality', we call the states \(|n\ket\) `beables', in honour of J.S. Bell\,\cite{Bell1987.ref}
The theory known as `quantum mechanics' is now nothing but the \emph{vector representation} of such evolution laws. This means that we regard the beables as elementary vectors that form the basis of a Hilbert space. It allows us to define \emph{energy} \(E\), by re-arranging the beables in eigenstates \(|k\ket_E\) of the evolution operator \(U_{\Del t}\):
	\be |n\ket=\frac 1{\sqrt N}\sum_{k=0}^{N-1}e^{-2\pi ikn/N}|k\ket_E\ , \qquad 
	|k\ket_E=\frac 1{\sqrt N}\sum_{n=0}^{N-1}e^{2\pi ikn/N}|n\ket\ ,  \eel{discFourier.eq}
Consequently, \be U_{\Del t}|k\ket_E&=&e^{-2\pi i k/N}|k\ket_E\ , \quad\hbox{so that}\\[3pt]
	\quad U_t|k\ket_E&=&e^{- i E t}|k\ket_E\ ,\quad\hbox{with}\quad E=2\pi k\,/\,N\Del t\ , \ee
where \(U_t\) describes the evolution over \(t/\Del t\) time steps. Note that the maximum energy is
\be E^\mmax=2\pi (1-1/N)\,/\,\Del t\ . \ee

We can define the Schr\"odinger equation as
	\be \frac\dd{\dd t}|k\ket_E=-\frac{2\pi i k}{N\Del t}\,|k\ket_E\ , \ee
and evaluating the operator \(E\) in the basis of the beables \(|n\ket\), using transformations \eqn{discFourier.eq}, is straightforward.

By regarding the entire theory of quantum mechanics as the vector representation of a deterministic system, one may conclude that there is absolutely nothing mysterious about quantum mechanics. In particular, \emph{any} discrete, realistic, deterministic, classical system can be regarded as a large set of periodic subunits, which will always allow us to set up  a Schr\"odinger equation along lines as sketched above. One may \emph{always} `postulate' Born's rule of regarding the absolute squares of amplitudes as probabilities, and one may always `postulate' the rule that a wave function collapses after a measurement, and explain the outcomes of experiments along such lines. Thus, what used to be distinct axioms of quantum mechanics, can actually be seen as man-made rules to interpret superpositions as probability distributions. All mysteries then evaporate.

To illustrate this, the author set up a generic program to mimic any Schr\"odinger equation, to any accuracy\cite{GtH-2020.ref, GtHLHV.ref}, by starting out with periodic variables that interact weakly (an interaction that is as real and deterministic as the periodic ones above). This program works by noting that variables that are periodic at the tiniest conceivable time scale, will always turn out to be in their lowest energy states, and this can be used to express all dynamics in terms of the energy eigen modes, leading to complete freedom to derive any Schr\"odinger equation one wants. An interesting aspect of this derivation is that some Schr\"odinger equations will be easier to obtain than others, and this might become a lead in future model building attempts.

The resulting theory for quantum mechanics as resulting from smearing the fastest variables shows strong similarities with Nelson's stochastic theory of quantum mechanics\,\cite{Nelson.ref}, although we believe to have more universal mathematical justifications and procedures.

There are, however, a number of questions that are still quite legitimate. One may ask: ``How come that J.S.~Bell arrived at very disturbing contradictions by using his theorem?"\cite{Bell1964.ref,Bell1987.ref,Bell1982.ref}. This question can be answered in various ways. One answer is that Bell assumed his observers Alice and Bob to have `free will', whereas the notion of free will cannot be defined along the lines above. Bell's \emph{actual} problem was his fear of `conspiracy', or superdeterminism: the laws of physics seem to conspire to overthrow his carefully derived theorem. Answer: we can choose our classical system as complicated as we want, but at the classical level there is never any conspiracy. Conspiracies appear to arise if one tries to write all physical states, including all their probabilities, as quantum superpositions with well defined phases. But these are man-made. The quantum superposition coefficients are not entirely observable, and consequently, they, and particularly the phases of the superposition coefficients, may seem to behave not quite logically. The classical system does not know about superpositions and phases. 

In view of this, we think the source of the contradiction Bell arrived at, is that he still described the atoms and photons as if the quantum states they were in should be, to some extent, observable. This contradiction disappears completely only if one uses \emph{nothing but} the language of the really observable beables.

And yet, questions still remain. One may ask: Bell's theorem, and numerous similar mysteries, arise in particular whenever we consider spin states that can be rotated. Rotation symmetry itself \emph{is} a mystery. How can discrete, finite systems such as fields on a lattice, display invariance under continuous symmetry groups? Now this is a hard question. It is hard to rotate beable states over fractions of angles. We could argue that continuous symmetries may be `emergent phenomena', but detailed, rigorous  mathematical investigations that elucidate these phenomena have not yet been performed. So we haven't completed our job. There is one observation concerning continuous rotation symmetries: we are blessed with computer programs that allow us to rotate any photograph any way we like, as if there were no pixels, which are certainly not rotation invariant. Thus we note that emergent, continuous symmetries perhaps should not worry us, they may well occur in nature, but a more precise mathematical treatment would be welcome.

 \end{document}